\documentclass[11pt]{article}
\usepackage{sustyle}
\usepackage{tikz}
\usetikzlibrary{arrows.meta}
\usetikzlibrary{arrows}
\tikzstyle{main node}=[draw]

\usepackage[T1]{fontenc}
\usepackage{amsmath, graphicx, tikz, enumerate, amssymb, pgf}
\usepackage[colorlinks=true, allcolors=black,citecolor=black, urlcolor=black]{hyperref}

\usepackage[round]{natbib}
\usepackage{bibentry}
\usepackage{setspace}
\usepackage{pgfplots}
\usepackage{bchart}

\usepackage{enumitem}
\usepackage{booktabs}

\newcommand\blfootnote[1]{%
  \begingroup
  \renewcommand\thefootnote{}\footnote{#1}%
  \addtocounter{footnote}{-1}%
  \endgroup
}

\title{\Large Do Large Language Models (Really) Need Statistical Foundations?}

\date{May 2025; Revised January 2026}

\begin{document}

\author{Weijie Su\\[0.5em]University of Pennsylvania}

\maketitle

\begin{abstract}
Large language models (LLMs) represent a new paradigm for processing unstructured data, with applications across an unprecedented range of domains. In this paper, we address, through two arguments, whether the development and application of LLMs would genuinely benefit from foundational contributions from the statistics discipline. First, we argue affirmatively, beginning with the observation that LLMs are inherently statistical models due to their profound data dependency and stochastic generation processes, where statistical insights are naturally essential for handling variability and uncertainty. Second, we argue that the persistent black-box nature of LLMs---stemming from their immense scale, architectural complexity, and development practices often prioritizing empirical performance over theoretical interpretability---renders closed-form or purely mechanistic analyses generally intractable, thereby necessitating statistical approaches due to their flexibility and often demonstrated effectiveness. To substantiate these arguments, the paper outlines several research areas---including alignment, watermarking, uncertainty quantification, evaluation, and data mixture optimization---where statistical methodologies are critically needed and are already beginning to make valuable contributions. We conclude with a discussion suggesting that statistical research concerning LLMs will likely form a diverse ``mosaic'' of specialized topics rather than deriving from a single unifying theory, and highlighting the importance of timely engagement by our statistics community in LLM research.

\end{abstract}

\blfootnote{Email: \texttt{suw@wharton.upenn.edu}.}

\section{Introduction}

Consider a thought experiment where an octopus under the seabed connects to a submarine cable and then eavesdrops on human conversations without any prior knowledge of human language. All it has access to are the utterances of one speaker and the corresponding responses of the other. One may ask: can the octopus ever learn to understand human language, the content of these conversations, and even possess some level of intelligence, purely based on human conversations by passively listening to arbitrarily large amounts of these paired observations?

This thought experiment illustrates how large language models (LLMs) are essentially developed \citep{kottke2023octopus}. LLMs are massive neural networks, predominantly based on the Transformer architecture \citep{vaswani2017attention}, trained on immense text corpora---including human-written content, code, and various other forms of text---to predict the next word (formally called a ``token'') given the preceding sequence of tokens \citep{radford2018improving,brown2020language}. By design, an LLM is an autoregressive model that attempts to learn language purely by learning statistical patterns in human-generated text, rather than explicit linguistic rules or semantic grounding.

Interestingly, when Ilya Sutskever and his colleagues at OpenAI proposed next-token prediction as a training strategy in 2018 \citep{radford2018improving}, very few researchers believed that such a simple training paradigm would yield capabilities resembling ``understanding'' of language \citep{bender2021stochastic,suprisentp}.\footnote{In this paper, we exclude non-GPT models such as BERT from our discussion, which use a masked-language-modeling objective \citep{devlin2019bert}.} In stark contrast, when browser-based ChatGPT launched in late 2022, it created a global sensation as the public marveled at its ability to generate human-like text and handle wide-ranging tasks \citep{bubeck2023sparks}.\footnote{Although the NLP community had already been impressed by GPT-3 in 2020, ChatGPT's web interface made these advances far more widely accessible.} Not only did ChatGPT impress users with its near Turing test level conversational abilities \citep{jones2025large}, but it also demonstrated elementary reasoning skills---enough to carry out basic statistical analyses and data visualization \citep{tu2024should,lin2025divergent}. These capabilities have continued to advance at a rapid pace.

Given that LLMs involve many statistical aspects, a burgeoning body of statistical research on LLMs has emerged (see a recent overview by \citet{ji2025overview}). While this review paper provides a broad, tutorial-style introduction for statisticians who wish to study LLMs and obtain a big-picture view, the present paper is a shorter, more focused and selective piece that concentrates on whether LLMs would genuinely benefit from the kinds of statistical foundations that our community has developed and could further develop, rather than offering a comprehensive review. Furthermore, we ask if such statistical contributions would lead to improved development, deployment, and application, particularly to guide and enhance their real-world use? In other words, we move beyond the discussion of how LLMs can be used to enhance statistical analysis and statistical education to focus on whether statistical methodology and insights can improve LLMs themselves.\footnote{For clarity, this paper does not address how LLMs can be used to enhance statistical analysis or statistical education. Interested readers are referred to \citet{tu2024should}.}

A straightforward argument supporting the need for statistics in LLMs comes from Richard Sutton's influential essay, \textit{The Bitter Lesson} \citep{sutton2019bitter}. The recent Turing Award laureate observed that enduring progress in artificial intelligence (AI) over decades has primarily stemmed from leveraging increased computational power and data scale, rather than incorporating hard-coded human knowledge or intricate model design. Statistical approaches are well suited to leveraging massive data via computation, as they are designed to let the data speak for themselves. Indeed, Sutton articulated in his essay that ``statistics and computation came to dominate the field'' of AI.

Given the repeated validation of the bitter lesson \citep{yousefi2024learning}, there are compelling reasons to believe that statistics will undoubtedly benefit LLMs. However, we should recognize that the term ``statistics'' in Sutton's context is broad. It encompasses both purely predictive algorithms---such as neural networks, boosting, and random forests---and what might be termed inferential statistics. This latter category roughly corresponds to what Bradley Efron termed ``estimation and attribution'' methods \citep{efron2020prediction} or what Leo Breiman called ``data-modeling'' methods in his \textit{Two Cultures} paper \citep{breiman2001statistical}. These approaches focus on interpretable parameters and uncertainty quantification to make inferences from noisy observations. In this paper, we ask whether LLMs, as they become more powerful and widely deployed, would benefit from the interpretable, inferential, and uncertainty-aware techniques that statistics in this narrower sense can offer.

We argue affirmatively that LLMs require more statistical contributions for their continued advancement. Although LLMs by design are purely predictive algorithms, they differ significantly from prior methods in this category, including even pre-Transformer neural networks \citep{krizhevsky2012imagenet,simonyan2014very}. This distinction emerges in two key ways: First, LLMs operate on an unprecedented diversity of data types, including almost all possible forms of text data. While earlier models could process text \citep{manning1999foundations}, it is the first time a single model can integrate and process text almost as seamlessly as numbers, enabling a unified approach to handling tasks as diverse as code generation, language translation, and data analysis, marking a substantial departure from previous predictive algorithms primarily focused on structured or image data. This strong data dependency positions LLMs as compressors of vast human data, as suggested by the influential metaphor ``ChatGPT is a blurry JPEG of the web'' \citep{chiang2023chatgpt}. Second, the generative and stochastic nature of LLMs, which arises from next-token prediction, makes the models themselves random, with outputs necessarily involving variability and uncertainty \citep{huang2025survey}. Together, these characteristics lead to numerous statistical challenges and opportunities for principled treatment of variability, uncertainty, calibration, and inference, especially when LLMs are used in high-stakes decision-making.

Moreover, we contend that for many LLM-related problems, statistics may be not just useful but sometimes perhaps the only viable approach. To make this case, we highlight the black-box status of LLMs, meaning it is difficult to understand how they arrive at their decisions. While the degree of black-boxness can vary, LLMs are arguably among the most complex digital systems ever constructed. We will provide evidence that this black-box nature is not merely a temporary state but likely to persist. Consequently, deriving LLM behavior from first principles via mathematical modeling, as is often possible in physics, appears highly challenging, if not infeasible. As such, when direct mathematical modeling is impractical, statistical modeling often offers a flexible and effective approach to shedding light on complex systems. Indeed, statistical approaches allow us to posit and test relationships between observable variables (like inputs and outputs) and potentially unobservable latent factors, even before their internal mechanisms are fully understood \citep{candes2020discussion}. Examples include understanding how different data mixtures influence model capabilities or quantifying the uncertainty of the outputs of LLMs using conformal prediction \citep{liu2024regmix,mohri2024language,cherian2024large}.

Having argued for the fundamental necessity of statistics for LLMs, we outline several statistical research avenues on LLMs, which are illustrative rather than exhaustive, reflecting the rapid evolution of the field. These range from the formulation of principled methods for aligning models with human values, to leveraging the models' probabilistic outputs for tasks like content verification, to rigorously evaluating model capabilities and quantifying output uncertainty, and finally to optimizing the central role of data in shaping LLM performance. Given that LLMs operate in an end-to-end manner---directly processing text inputs to produce text outputs that can inform decisions or trigger actions---they permeate an ever-wider range of applications, continuously generating new statistical challenges and opportunities. Echoing Efron's sentiment about algorithmic developments originating outside statistics, ``future progress, especially in scientific applicability, will depend heavily on us'' \citep{efron2020prediction}.

\section{LLMs as Statistical Models}
\label{sec:statistical}

While designed as a predictive method where the underlying data-generating mechanism is treated as unknown, LLMs exhibit characteristics that distinguish them significantly from many other predictive algorithms \citep{breiman2001statistical,efron2020prediction}. Predictive methods like decision trees or support vector machines often rely on feature engineering or, equivalently, carefully chosen kernels, whereas modern LLMs operate in a nearly end-to-end data-centric manner, trained directly on vast quantities of raw data with minimal human intervention beyond data selection and cleaning.\footnote{Indeed, data curation is nontrivial in practice, but from a methodological perspective, no detailed human knowledge of linguistic constructs is explicitly hard-coded into LLMs at training time.} Nevertheless, human intervention remains significant regarding the choice of architectural designs for LLMs.

Owing to this strong data dependency, the capabilities of LLMs are largely determined by the properties and scale of the training data \citep{thrush2024improving}. This fact is quantitatively captured by scaling laws, which demonstrate a predictable relationship between model capabilities and the volume of training data during the pre-training phase \citep{kaplan2020scaling}. In essence, scaling laws suggest that LLMs are what they are trained on, elevating data to the most critical component of their development. As evidence, while open-source LLM developers often release model weights and sometimes some technical details of their training processes, the composition and preparation of their training datasets are almost never disclosed \citep{grattafiori2024llama,liu2024deepseek}.

The pivotal role of data continues after the pre-training phase. To achieve proficiency in complex tasks, LLMs typically undergo specialized post-training \citep{ouyang2022training}. This necessitates the use of vast corpora of high-quality and meticulously annotated examples to impart nuanced understanding and desired behaviors to LLMs. This has led to the rise of a substantial data-labeling and annotation industry for training, refining, and aligning AI models.

One might argue that this data-centricity is shared to some extent with smaller deep learning models, not solely LLMs. Yet LLMs stand apart due to two characteristics:

\paragraph{Anything as numeric.} LLMs operate directly on unstructured information---plain language, code, numbers, or even symbolic mathematics---by converting diverse data types into high-dimensional numeric vectors often thought to lie in a ``semantic'' space \citep{mikolov2013distributed}. This enables flexible execution of transformations within this semantic space, which LLMs ultimately map back to text. This allows LLMs to process anything representable as text almost as readily as regression models handle numbers. In effect, LLMs instantiate a general-purpose engine operating on numeric representations of virtually all forms of text data. Indeed, Geoffrey Hinton metaphorically described GPT-4 as emerging like a butterfly from billions of nuggets of understanding accumulated throughout human history \citep{hinton2023tweet}.

\paragraph{Stochastic nature of generation.} The predominant training paradigm for modern LLMs is next-token prediction \citep{radford2018improving}.\footnote{While other objectives exist (e.g., masked language modeling in BERT) and new approaches are emerging (e.g., diffusion models for text), the core task remains generative.} Crucially, this generative process is inherently stochastic, depending not only on the prior context but also on the data used for training LLMs. This randomness is not merely an artifact but arguably a necessity for modeling human language, which itself---whether in stories, manuals, essays, creative writing, code, or mathematical proofs---is generative and rarely follows a deterministic path. This contrasts sharply with many earlier classical neural networks, particularly for classification, where outputs often correspond to fixed, deterministic ground truths.

These two characteristics---the universal processing of data through numerical embedding and the inherent stochasticity in generation---arguably position LLMs in a way that resonates more closely with data-modeling or inferential methods than typical predictive algorithms. Recognizing this allows us to appreciate the potential and necessity of statistical insights for LLMs beyond their initial purely predictive design. Consequently, we can effectively treat LLMs as \textit{statistical models}.

Statistics becomes particularly relevant when a system involves dependence on data and the inference of patterns from that data. Indeed, the data-hungry nature of LLMs, coupled with their ability to process ``anything as numeric,'' leads to complex data-dependent patterns. The sheer scale and heterogeneity of LLM training data present many opportunities and significant challenges from a statistical perspective. For instance, understanding how different data sources contribute to specific model capabilities (e.g., models trained on larger proportions of code tend to empirically exhibit stronger programming abilities) is crucial for optimizing data mixtures to achieve desired performance, in both the pre-training and post-training phases \citep{xie2023doremi}. Furthermore, as suggested by researchers like Ilya Sutskever, relying solely on existing human-generated data may already be insufficient \citep{sutskever2024sequence}, and therefore the generation of high-quality synthetic data becomes vital. This is a task where statistical principles for experimental design and data augmentation could prove invaluable.

Furthermore, the variability and uncertainty stemming from the stochastic nature of LLMs demand statistical analysis. Because LLM outputs are inherently variable, understanding and quantifying the uncertainty associated with their responses is critical. This is especially true in scientific applications where reproducibility is required, and in high-stakes decision-making, such as medical diagnostics, where the model's confidence level can drastically influence subsequent actions. Statistics offers a rich toolkit for analyzing variability, uncertainty, and miscalibration, and subsequently reducing them. Conversely, recent work demonstrated that LLM stochasticity can be harnessed to construct statistically valid inference procedures \citep{ratkovic2025large}.

Conversely, the stochastic generation process not only presents challenges but also enables novel statistical techniques. Watermarking, for instance, leverages the model's probabilistic token generation to embed statistically detectable signals based on pseudorandomness, allowing for provable distinction between AI-generated and human-written text, potentially without compromising quality \citep{kirchenbauer2023watermark}. Another example is LLM alignment via reinforcement learning from human feedback (RLHF) \citep{ouyang2022training}, which uses the Bradley--Terry model \citep{bradley1952rank} to represent LLMs' preference distributions. Such techniques would be impossible if LLM generation were deterministic.

The need for statistical foundations is further amplified by the unprecedented breadth of LLM applications. LLMs are being integrated into coding assistants, writing tools, autonomous agents, scientific discovery platforms, medical information retrieval, and countless other domains. This rapid and widespread deployment into diverse, often novel environments constantly surfaces new challenges related to privacy, copyright, attribution, fairness, ethical considerations, and the need for mechanisms like machine unlearning \citep{cao2015towards} to remove specific sensitive knowledge. While addressing these complex issues for the trustworthy use of LLMs undoubtedly requires interdisciplinary collaboration, statistical insights and methodologies are particularly crucial and effective in developing solutions, with the additional advantage of typically being computationally light.

\section{LLMs as Black-box Models}
\label{sec:black}

The data-centric and stochastic characteristics of LLMs already present a compelling case for incorporating statistical methodologies into their development and application. Nevertheless, one might wonder whether non-statistical approaches---for instance, purely mechanistic, first-principles methods---could also achieve similar goals. In this section, we deepen the argument by positing that for many crucial challenges surrounding LLMs, statistics may not merely be useful but often the most viable, or perhaps the only practical, path forward. This necessity arises primarily from the profound \textit{black-box} nature of LLMs.

Whether a field leans heavily on statistical methodology often depends on the extent to which its underlying mechanisms are understood. Fields like classical physics, where fundamental principles are generally well-established, can often rely on deductive mathematical modeling to predict system behavior. While empirical data remains crucial for validation, data analysis often serves to confirm theories or estimate parameters within known models. Conversely, in fields like biology, particularly neuroscience, a vast number of unknowns and high-dimensional interactions render the internal workings inaccessible or too complex to model from first principles. Consequently, when confronting these kinds of black-box situations, researchers typically rely heavily on statistical inference to discern patterns, test hypotheses, and build useful predictive or explanatory models directly from observational data, even without a complete mechanistic understanding. It is arguably this difference in mechanism transparency, among other factors, that contributes to the reality that statistical methodologies are more prevalent in biomedical research than in physics.

We argue that LLMs, in their current state and likely trajectory, fall firmly into the category of complex systems where the black-box nature necessitates a reliance on statistical approaches. Their status as black boxes is likely a persistent feature arising from the following two factors:

\paragraph{Inherent complexity and huge scale.} LLMs are among the most complex computational systems ever built. The Transformer architecture, upon which almost all proprietary LLMs are based, involves intricate compositions of a variety of components such as multi-head attention, layer normalization, gating mechanisms, and nonlinear transformations, interacting across billions to nearly trillions of parameters \citep{vaswani2017attention}. Indeed, the sheer size of LLMs appears necessary. On the theoretical front, \citet{bubeck2021universal} showed that neural networks satisfying certain regularity conditions necessitate a vast number of parameters. Empirically, scaling laws further confirm that performance consistently improves with model size \citep{kaplan2020scaling}. This immense scale makes a detailed, analytical understanding from bottom-up principles practically intractable.

\paragraph{Non-uniqueness of architectures and optimizers.} The black-box nature is compounded by the fact that there is not one single ``correct'' architecture for LLMs achieving high performance. While the Transformer has been dominant, simplified Transformer and mixture-of-experts Transformer variants are shown empirically to work well \citep{he2024simplifying,dai2024deepseekmoe}.\footnote{Notably, the Transformer can learn not only data patterns but also entire algorithms for statistical problems \citep{he2025learning,cao2025transformers}.} Moreover, non-attention-based architectures like state-space models (Mamba) \citep{gumamba}, recurrent structures (RWKV) \citep{peng2023rwkv}, and even potentially large-scale LSTMs with sufficient memory show promise or competitive results \citep{Schmidt2023Transformers}. Similarly, various optimization algorithms---including Adam \citep{kingma2014adam}, stochastic gradient descent (SGD) \citep{robbins1951stochastic}, AdamW \citep{loshchilov2017decoupled}, Shampoo \citep{gupta2018shampoo}, and the more recent Muon optimizer \citep{Keller2024Muon,lau2025polargrad,su2025isotropic}---have been effective in training these massive models. This lack of a uniquely optimal design for LLMs is reflective of the empirical trial-and-error approach driving both architectural and algorithmic innovations. Indeed, this empirical flexibility fosters co-adaptation between architectures and optimizers. Interestingly, a popular hypothesis within the AI community posits that Adam's effectiveness may partially result from neural architectures being inadvertently ``overfitted'' to its optimization characteristics \citep{orabona2020neural}.

The confluence of immense complexity, necessarily large scale, and non-unique design makes it highly improbable that we can understand the behavior of LLMs from neatly closed-form laws in the way physicists often model physical phenomena. LLMs are thus de facto black boxes and likely to remain so for the foreseeable future (see Section~\ref{sec:discussion} for elaboration). As Stephen Wolfram has argued, complex systems may sometimes be computationally irreducible, meaning their behavior cannot be predicted by simple, interpretable rules \citep{wolfram2003new}. Consequently, attempts to build comprehensive mathematical theories to unveil the inner workings of LLMs are likely to face fundamental challenges, if not prove impossible in practice.\footnote{For completeness, we recognize efforts
  in mechanistic interpretability that aim to shed light on internal computations. These investigations, however, are computationally intensive and often yield localized or partial insights rather than a comprehensive, predictive approach to demystifying the entire system \citep{elhage2021mathematical}.} Furthermore, the complex lifecycle of LLMs \citep{alaa2024veridical}, often involving opaque upstream development processes and limited transparency for downstream users, inherently challenges general efforts toward transparency.

When dealing with a complex, stochastic black-box system where the true underlying process is unknown or intractable, we must resort to studying the system through its inputs and outputs, along with any available intermediate measurements and potentially latent factors. This necessitates the use of approximate and data-driven models to capture observed behaviors. Such models, built from data to approximate an unknown or intractable underlying process, are inherently statistical. The resulting statistical models are, in the spirit of George Box, necessarily wrong in the sense of being incomplete, yet potentially useful for prediction, understanding correlations, and guiding development \citep{box1976science}. Indeed, this perspective resonates with Alexei Efros' call to treat AI research more like experimental biology, focusing on using statistical methodology to make progress based on empirical observation, hypothesis testing, and data-driven modeling \citep{Efros2023}.

\section{Statistical Topics on LLMs}
\label{sec:exampl-stat-topics}

In this section, we illustrate a number of research directions where we believe statistical principles can directly enhance both the development and application of LLMs. These directions either exploit the generative and data-driven essence of LLMs (Section \ref{sec:statistical}) or take the viewpoint of LLMs as complex black-box systems (Section \ref{sec:black}). This list is not exhaustive, reflecting the dynamic nature of the field, but serves to highlight the breadth of opportunities, and we anticipate that additional statistical challenges will arise as LLM capabilities and deployment settings continue to evolve. Moreover, many of these research areas demand modest computational resources, often requiring only API access to existing models, thus making them accessible to many researchers.

\paragraph{LLM alignment.} Alignment is the process of steering AI models toward human preferences, intended goals, and ethical principles. Because human preferences and ethics can often be represented quantitatively via statistical models, statistical principles naturally arise for developing principled and trustworthy alignment methods.

\begin{itemize}
    \item \textit{Alignment from human feedback:} The technique of RLHF involves training a reward model based on human comparisons of LLM outputs \citep{ouyang2022training}. Formally, the preference distribution between two possible responses $y, y'$ to a prompt $x$ is modeled using the Bradley--Terry model:
\[
\mathbb{P}(y \text{ is preferred over } y'|x) = \frac{\e^{r(x, y)}}{\e^{r(x, y)} + \e^{r(x, y')}},
\]
where the reward $r(x, y)$ is trained from pairwise comparison data from human labelers using maximum likelihood estimation. The LLM is then fine-tuned using the reward model to maximize the expected reward, subject to a constraint that penalizes excessive deviation from the reference model. Owing to the stochasticity inherent in the Bradley--Terry model with noisy human feedback, this process is statistical in nature. This offers numerous opportunities for statisticians to analyze reference model misspecification, sample efficiency of preference data collection, and the generalization of learned preferences \citep{zhu2023principled,chakraborty2024maxmin,swamy2025all,ye2025robust,fengpilaf}. For example, recent studies demonstrate that the current approaches to RLHF inadequately represent the full spectrum of human preferences and can introduce statistical bias during fine-tuning \citep{liu2025statistical, xiao2024algorithmic}. Another notable direction is to leverage experimental design for optimizing data collection for alignment, which is often costly. These approaches include adaptive data collection \citep{mukherjee2025adaptive}, optimal design for reward modeling \citep{scheid2024optimal}, and selection of sampling strategies \citep{shicrucial}.

\item \textit{Privacy and machine unlearning:} LLMs trained on vast datasets may inadvertently memorize and expose sensitive or copyrighted information. Differential privacy provides a rigorous mathematical framework that offers statistical guarantees against information leakage by introducing controlled noise during training \citep{dwork2006calibrating}. A critical statistical consideration, therefore, is to optimize the trade-off between privacy protection and model utility, particularly for proprietary LLMs. Recent research has focused on enhancing this trade-off by implementing differential privacy during the fine-tuning phase of pre-trained models that were initially trained on public data \citep{lilarge2022} and generating differentially private synthetic data using LLMs without requiring model training \citep{xie2024differentially}. However, further research is needed to meet the level of trade-off required by proprietary LLMs. A related area is machine unlearning, which aims to efficiently eliminate the influence of specific data points (e.g., due to privacy requests or copyright concerns) without retraining \citep{cao2015towards}. This approach presents significant statistical challenges in precisely defining and verifying effective ``forgetting'' while preserving model capabilities \citep{yao2024large,zhangnegative}.

\item \textit{Fairness:} LLMs can inherit and amplify societal biases present in their training data, leading to discriminatory outputs across different demographic groups \citep{santurkar2023whose,kotek2023gender}. While addressing fairness is a complex socio-technical problem, statistics provides indispensable tools for defining, measuring, and mitigating bias. Recent empirical research has moved beyond simple disparity metrics to statistical frameworks that account for the stochastic nature of generation. For instance, \citet{liu2024bias} introduced a bias-volatility decomposition which reveals that alignment techniques like RLHF often reduce systematic bias at the cost of increasing volatility, implying that the model's behavior becomes statistically less predictable for minority groups. Furthermore, evaluations of mitigation strategies indicate that surface-level alignment often fails to remove underlying dependencies \citep{bai2025explicitly}. In light of this, open statistical problems remain, particularly in developing sample-efficient auditing procedures for subgroups, formulating fairness frameworks to distinguish between stereotypes and necessary semantic correlations, and incorporating fairness considerations into various stages of the LLM pipeline---from data curation and pre-training objectives to alignment processes or even directly during generation of outputs \citep{zhang2024fair,chakraborty2024maxmin}.

\end{itemize}

\paragraph{Exploiting the generative interface.} The autoregressive nature of LLMs, specifically, next-token prediction, allows one to treat LLMs as a black-box machine that outputs a multinomial distribution, which is used to sample the next token. This probabilistic viewpoint allows us to develop statistical methods by leveraging the sampling properties of multinomial distributions, without needing to delve into the complexity of how the Transformer architecture computes the distributions.

\begin{itemize}
    \item \textit{Watermarking:} To distinguish LLM-generated text from human-written content, watermarking techniques embed statistically detectable signals into the generation process based on cryptographic pseudorandomness \citep{kirchenbauer2023watermark,scott2023watermarking}. Formally, the next token $w_{t+1}$ is decoded as $\mathcal{S}(\bm P_t, \zeta_t)$, where the decoder $\mathcal{S}$ is deterministic or can incorporate sampling randomness, $\bm P_t$ is the multinomial distribution for drawing the $(t+1)^\textnormal{st}$ token, and $\zeta_t$ is a pseudorandom variable that can be computed from the preceding context and a private key. The detection problem can be naturally framed as statistical hypothesis testing, based on the observation that when text is not watermarked (under the null hypothesis), $w_{t+1}$ is independent of $\mathcal{S}(\bm P_t, \zeta_t)$, while when the text is watermarked (under the alternative hypothesis), $w_{t+1} = \mathcal{S}(\bm P_t, \zeta_t)$. The latter case necessarily induces dependence between the tokens and pseudorandom variables, even without knowledge of the multinomial distributions. This framework opens avenues for applying statistical decision theory to design watermarking schemes and detection rules that achieve optimal or near-optimal detection performance while preserving text quality \citep{li2025statistical}. A significant practical challenge involves ensuring robustness against adversarial modifications, such as paraphrasing or translation, necessitating the development of schemes with provable statistical guarantees under such attacks \citep{pang2024attacking,li2024robustdetectionwatermarkslarge}. Additionally, there is substantial interest in leveraging watermarking techniques for other purposes, such as detecting data misappropriation \citep{cai2025statistical}.

    \item \textit{Speculative Sampling:} This technique accelerates text generation of LLMs by employing a smaller, faster ``draft'' model to propose candidate tokens, which are subsequently accepted or rejected by the larger ``target'' model based on a comparison of their respective output distributions \citep{leviathan2023fast, chen2023accelerating}. To add detail on speculative sampling, let $\bm P_t$ and $\bm Q_t$ denote the multinomial distributions of the larger and smaller LLMs, respectively, for predicting the $t^\textnormal{th}$ token---assuming the index of the last token in the prefix is 0. Denote by $x_1, \ldots, x_T$ the tokens sequentially proposed by the smaller LLM conditioned on the preceding tokens. Then, beginning from $t=1$, the larger LLM accepts token $x_t$ with probability $\min\{1, P_t(x_t)/Q_t(x_t)\}$, and terminates the process when the first rejection occurs. The final step of speculative sampling appends one additional token for this epoch by sampling from $\max\{\bm 0, \bm P_t - \bm Q_t\}$ after normalization, where $t$ denotes the position at which the smaller LLM's proposed token is first rejected. As is clear, speculative sampling is a form of rejection sampling that achieves the maximum coupling between the prediction probabilities of the two LLMs.

The rationale behind this technique is that, for the larger LLM, evaluating the probability of a proposed token is computationally more efficient than decoding a token, which typically requires processing all possible tokens in the (extensively large) vocabulary. Consequently, the efficiency gain depends critically on the acceptance rate, which is inherently a statistical quantity depending on how the target LLM's probability distributions align with those of the smaller one. Statistical analysis is crucial for optimizing the trade-off between the computational cost of the draft model and the expected speedup from the acceptance rate. For instance, this statistical technique has been implemented in DeepSeek V3 for efficient multi-token prediction \citep{li2024eagle,liu2024deepseek} and can be integrated with watermarking \citep{hu2024inevitable}. More opportunities exist for incorporating statistical insights into adaptive variations of this technique and integrating it with other methods.

    \item \textit{Tokenization:} Tokenization breaks down text into discrete tokens to form the categories from which LLMs sample. Tokenization fundamentally impacts the statistical
properties of the data fed into the model and the distributions it outputs. However, most current tokenizers (e.g., byte-pair encoding \citep{gage1994new}) are based on heuristic compression algorithms and lack statistical guarantees. There is a need for statistically principled tokenization methods that optimize for criteria like information rate or minimal sequence length across diverse text types. Furthermore, statistical analysis is required to understand how tokenization efficiency and potential induced biases vary across different languages, domains (e.g., code, scientific literature), and demographic groups \citep{phan2024understanding, yang2024problematic}.
\end{itemize}

\paragraph{Assessment of LLM behavior.} Understanding and quantifying the reliability, limitations, and capabilities of  LLMs presents significant challenges, exacerbated by their stochastic and black-box nature. This inherently necessitates statistical modeling for assessing LLM behavior with confidence statements.

\begin{itemize}
    \item \textit{Uncertainty quantification and calibration:} LLM outputs exhibit uncertainty stemming from both inherent randomness in the generation process and knowledge limitations \citep{yadkori2024believe}. For trustworthy applications, particularly in high-stakes scenarios, quantifying the uncertainty of LLM outputs becomes crucial. Among various approaches addressing uncertainty, conformal prediction has emerged as a statistically rigorous and flexible methodology for providing prediction sets with distribution-free coverage guarantees---a characteristic particularly suited to the black-box nature of LLMs \citep{mohri2024language,cherian2024large,guiconformal}. Furthermore, the utility of uncertainty estimates depends on calibration---the degree to which a model's expressed confidence aligns with its actual accuracy. Given that aligned LLMs are often miscalibrated \citep{achiam2023gpt}, developing methods that simultaneously quantify uncertainty and restore calibration is an important research direction \citep{xiao2025restoring,wang2025sconu,zhang2025cotuq}.

    \item \textit{Evaluation:} Assessing LLM capabilities across diverse tasks using benchmarks such as MMLU \citep{hendrycksmeasuring}, TruthfulQA \citep{lin2022truthfulqa}, and GSM8K \citep{cobbe2021training} is essential not only for tracking progress but also for guiding AI development \citep{silver2025welcome, yao2025secondhalf,gao2025comparison}. However, the probabilistic and complex nature of LLMs introduces substantial statistical challenges in evaluation. Statistically grounded methods are required to quantify the variance and reliability of evaluation scores \citep{miller2024adding, polo2024tinybenchmarks,li2025evaluating}, with statistical models such as item response theory being employed for this purpose \citep{madaan2024quantifying}. Nevertheless, this area faces an ``evaluation crisis,'' where reported benchmark scores frequently inflate perceived capabilities, partly due to evaluation gaming---the process of optimizing models specifically for benchmark performance \citep{khomenko2025}. This phenomenon bears resemblance to statistical issues such as $p$-hacking. Consequently, we need rigorous statistical principles for robust measurement and protection against overfitting to evaluation datasets. Techniques from experimental design can also enhance evaluation efficiency, for instance, by adaptively selecting test questions that best discriminate between model capabilities rather than using fixed benchmarks \citep{collins2024evaluating}.

\item \textit{Hallucination:} LLMs often generate fluent but factually unsupported content, a phenomenon referred to as hallucination. This is widely viewed as a ``last mile'' obstacle for deploying LLMs in high-stakes applications. Due to its critical importance, emerging work has been developed to detect hallucinations \citep{sriramanan2024llm,duan2024llms}, estimate hallucination rates \citep{jesson2024estimating}, and mitigate their occurrence \citep{huang2025survey}. Notably, \citet{kalai2024calibrated} identified a fundamentally statistical cause of hallucination: under a natural calibration condition and a Good--Turing-style approach to modeling facts, there is an inherent lower bound on the hallucination probability that does not depend on architecture or data quality \citep{kalai2025language}. While insightful, their probabilistic models remain highly stylized relative to the complexity of hallucinations observed in practice. Developing sequential probabilistic models over facts and prompts may provide a promising direction for bridging the gap between these theoretical results and the empirical behavior of hallucinations.

\end{itemize}

\paragraph{Central role of data.} The capabilities of LLMs are fundamentally determined by the data used in pre-training and fine-tuning \citep{yue2025does}. This gives rise to numerous statistical challenges related to understanding and optimizing the relationship between data characteristics and model performance.

\begin{itemize}
    \item \textit{Data mixture and attribution:} An essential challenge is determining the optimal composition of diverse data sources (e.g., web text, books, code, scientific papers) to train an LLM that achieves specific desired capabilities, often under resource constraints \citep{xie2023doremi}. While heuristic understanding exists---for instance, that a higher proportion of code in the training data generally leads to stronger coding abilities---the complex, high-dimensional relationship between data mixture and emergent abilities is largely unknown \citep{thrush2024improving}. Statistical modeling, particularly regression-based approaches, offers a simple yet effective approach to investigating these dependencies \citep{liu2024regmix}. Closely related to data mixture is the problem of data attribution, which seeks to identify the specific training samples that most influence a particular model output or behavior. Data attribution is crucial for addressing legal concerns like copyright infringement and enhancing transparency, but poses significant challenges due to the black-box nature of LLMs. Statistical techniques such as influence functions \citep{koh2017understanding} and kernel approximation \citep{park2023trak} have been used to address these challenges for relatively small-scale models, yet significant research effort is needed to adapt and scale these methods effectively for large models \citep{deng2024texttt,pan2025daunce}.

    \item \textit{Synthetic data and model collapse:} As the scale of LLMs continues to grow, relying solely on existing human-generated data (``the fossil fuel of AI,'' as put by Ilya Sutskever in his keynote speech at NeurIPS 2024) may become insufficient or cost-prohibitive. Consequently, synthetic data is becoming increasingly vital for its scalability and cost-effectiveness \citep{eldan2023tinystories, adler2024nemotron,yang2024synthetic,tian2025conditional}. Statistics offers valuable tools for guiding the synthetic data generation process, including methods for assessing data quality, controlling distributional properties to match desired targets, and designing efficient data augmentation strategies \citep{angelopoulos2023prediction}. However, a major risk arises from recursively training models on their own synthetic outputs, which can lead to a degradation of model quality, loss of diversity, and divergence from the true data distribution---a phenomenon termed model collapse \citep{shumailov2023curse}. Understanding the underlying mechanisms driving model collapse and developing statistically sound methodologies to mitigate it, potentially by adaptively mixing real and synthetic data or by imposing specific distributional constraints, represents a promising research direction where statistical insights are needed \citep{gerstgrasser2024model,dey2024universality}.

    \item \textit{Scaling laws:} Scaling laws are empirical observations that quantitatively relate an LLM's performance to factors such as the size of the training dataset, the number of model parameters, and the computational resources allocated for training \citep{kaplan2020scaling}. Among the various forms of scaling laws, \citet{hoffmann2022training} demonstrated that
\[
L = E + \frac{A}{N^{\alpha}} + \frac{B}{D^{\beta}}
\]
effectively captures how the pre-training loss $L$ depends on the number $N$ of model parameters and the number $D$ of training tokens, where $E$ denotes the entropy of natural text and $A, \alpha, B$, and $\beta$ are constants. These laws offer significant practical value by enabling researchers to predict potential performance gains from increased scale, thereby guiding model design, optimal model sizing, and resource strategies for training progressively larger models without requiring exhaustive experimentation \citep{achiam2023gpt}. Scaling laws are fundamentally statistical, as they model empirically observed relationships between the scale of training resources and model performance. From a theoretical perspective, however, scaling laws present intriguing questions for statisticians. The observation that model performance continues to improve with increasing model size $N$, seemingly without saturation even when $N$ becomes extremely large, challenges classical statistical learning theories that predict model performance saturation once complexity exceeds the intrinsic dimensionality of the data. Investigating the statistical underpinnings of these empirical laws, potentially through the lens of nonparametric estimation or approximation theory in high dimensions, may yield deeper insights into the learning dynamics of LLMs and potentially lead to novel statistical methodologies.

\end{itemize}

\paragraph{Summary.} The research topics discussed above are summarized in Table \ref{tab:stat_topics}. Broadly speaking, these directions fall into two complementary categories. The first is foundational and diagnostic, such as uncertainty quantification, hallucination, and evaluation. The second category is efficiency-driven and engineering-oriented, addressing practical needs such as data mixture optimization, watermarking, and synthetic data generation.

\begin{table}[!t]
\centering
\small % Reduces font size for the entire table
\caption{Summary of statistical research directions for LLMs. The topics are grouped by their section in the text.}
\label{tab:stat_topics}
\renewcommand{\arraystretch}{1.2}
% Adjusted column widths to fit the content better with small font
\begin{tabular}{@{}p{0.30\textwidth} p{0.24\textwidth} p{0.42\textwidth}@{}}
\toprule
Topic & Primary objective & Statistical methodology \\ \midrule
\multicolumn{3}{@{}l}{\textit{LLM alignment}} \\
Alignment from human feedback & Preference/Control & Bradley--Terry model, experimental design \\
Privacy \& machine unlearning & Security/Privacy & Differential privacy, forgetting verification \\
Fairness & Audit/Interpretability & Fairness metrics \\ \midrule
\multicolumn{3}{@{}l}{\textit{Exploiting the generative interface}} \\
Watermarking & Authentication/Detection & Hypothesis testing, decision theory \\
Speculative sampling & Efficiency/Speed & Rejection sampling, coupling \\
Tokenization & Efficiency/Compression & Information rate, compression algorithms \\ \midrule
\multicolumn{3}{@{}l}{\textit{Assessment of LLM behavior}} \\
Uncertainty quantification \& calibration & Reliability/Diagnostics & Conformal prediction, calibration \\
Evaluation & Measurement & Item response theory, experimental design \\
Hallucination & Reliability/Factuality & Good--Turing estimators, sequential models \\ \midrule
\multicolumn{3}{@{}l}{\textit{Central role of data}} \\
Data mixture \& attribution & Composition/Transparency & Regression, influence functions \\
Synthetic data \& model collapse & Performance/Stability & Sampling theory, distributional control \\
Scaling laws & Forecasting & Nonparametric estimation, approximation theory \\ \bottomrule
\end{tabular}
\end{table}

\paragraph{Other research directions.} The rapidly evolving landscape of LLMs continues to generate novel statistical challenges that extend beyond the aforementioned categories, many of which remain incompletely formulated from a statistical perspective and thus present opportunities for contributions from our community. In the development of small LLMs, which are needed for deployment in edge devices, empirical evidence has revealed that knowledge distillation from larger LLMs often outperforms training from scratch \citep{guo2025deepseek}, which calls for developing statistically efficient distillation methods. Accordingly, there is a need for owners of proprietary LLMs to develop sampling strategies that limit the effectiveness of distillation by business competitors \citep{savani2025antidistillation}. The recent emergence of reasoning models that employ latent intermediate steps via chain-of-thought \citep{wei2022chain} offers an effective approach to improving the reasoning abilities of LLMs \citep{muennighoff2025s1}. The promise of reasoning models calls for statistical methodology development in quantifying confidence for deciding when to terminate reasoning to save tokens while preserving accuracy \citep{sui2025stopoverthinking,zhang2025cotuq}, as well as for performing statistical inference on the resulting reasoning outcomes, including uncertainty quantification and the assessment of internal consistency across multiple reasoning paths \citep{zhang2025cotuq,wang2022selfconsistency}. Furthermore, the very recent advent of diffusion-based LLMs presents an opportunity for statistical analysis to elucidate the fundamental comparisons between the autoregressive and diffusion-based strategies \citep{nie2025large,dou2024optimal}. Moreover, as LLMs are frequently deployed as evolving API-based services, the development of statistically grounded techniques to detect unannounced updates and consequent behavioral shifts is critical for ensuring the reliability of downstream applications \citep{dima2025you}. Finally, the modification of multinomial distributions for next-token prediction through a Bayesian perspective represents another frontier where statistical insights can directly inform LLM development \citep{zhuang2025text}.

\section{Discussion}
\label{sec:discussion}

The past decade and a half, starting with the advent of AlexNet, has witnessed a remarkable advancement of purely predictive algorithms, with LLMs emerging as perhaps the most striking example. While LLMs share many defining aspects of purely predictive algorithms, their versatility in handling a variety of data types---especially unstructured text---and flexibility in applications across unprecedentedly diverse domains clearly distinguish LLMs from earlier purely predictive models, including pre-Transformer neural networks used in classification. Indeed, it may be more accurate to consider LLMs as enabling a new data-processing paradigm that converts and unifies diverse text-based inputs into a numeric form that can then be transformed and generated back into text, creating new forms of data amenable to analysis. Analogous to how genome-wide association studies once catalyzed the development of high-dimensional statistics, there are good reasons to believe that the continued progress of LLMs will open up an entire class of problems for which statistical methodologies will thrive.

Even taking a narrower scope, we argue further that classical inferential statistical principles---particularly those aligned with the ``data-modeling'' culture \citep{breiman2001statistical} and the ``estimation and attribution'' perspective \citep{efron2020prediction}---are becoming increasingly relevant for LLMs. The inherently stochastic nature of LLM generation makes statistical approaches suitable for quantifying and understanding uncertainty and variability. Moreover, while their architecture is in principle known, the stronger case for a statistical viewpoint emerges because the nearly trillions of parameters that LLMs operate on lack straightforward interpretation. Indeed, when faced with a system exhibiting black-box complexity, approximating its behavior through data-driven,
testable, and refutable statistical models is perhaps the only tractable and effective approach, as statistical methods have a long history of proving effective even when underlying mechanisms are not fully understood. This black-box complexity requires that we formulate solutions for some use cases of LLMs---especially those requiring stability, robustness, and interpretability---using smaller, interpretable probabilistic models or inferring relationships between interpretable factors. Indeed, current (and most likely future) approaches---whether mitigating alignment biases, verifying content origin, quantifying the reliability of generated content, or assessing the influence of specific data subsets---often involve statistical reasoning, hypothesis testing, or parameter estimation within a relevant probabilistic framework.

One may hope that the internal workings of LLMs will eventually become transparent and amenable to purely mechanistic analysis \citep{elhage2021mathematical}, which would perhaps eliminate the need for statistical approaches to the problems we present in Section~\ref{sec:exampl-stat-topics}. Yet strong evidence suggests otherwise.

\paragraph{Hypothesis of perpetual black-box state-of-the-art models.} A crucial argument supporting our claim is that state-of-the-art models are constantly evolving, driven by empirical gains achieved through increasing scale, architectural modifications often optimized for hardware efficiency, and new training heuristics. Consequently, the theoretical understanding of these models lags significantly behind practitioner-led advancements, a trend observable since the introduction of AlexNet \citep{su2024envisioning}, leading to a widening gap between what can be rigorously understood and the capabilities of the latest LLMs.

The persistent black-box status of LLMs suggests that a single, unifying ``grand statistical foundation for LLMs'' is unlikely to emerge. Consequently, statistical research in this area will likely proceed in a bottom-up fashion, driven by the need to solve specific problems and address particular applications. This problem-driven approach is expected to yield a mosaic of specialized statistical frameworks and techniques tailored to distinct challenges. While this implies diversity in methodology, our personal experience working on several LLM-related problems shows that fundamental statistical thinking and inferential reasoning remain consistently crucial and effective across different contexts. This lack of a single unifying foundation presents a wealth of research opportunities for statisticians with varied skillsets.

Although this paper primarily focuses on inferential or classical statistics, this emphasis does not imply that statistical research on LLMs should be exclusively inferential. Rather, progress will likely require a blend of both inferential and predictive statistical approaches. Indeed, the boundary between these two statistical cultures is often blurry, especially in practice. Furthermore, advancing statistical methodology in LLM research requires recognizing the interplay between statistics and data science, particularly the significant engineering component inherent in the latter \citep{donoho201750}. Embracing data science practices, such as robust data and code sharing, will also be vital for accelerating progress across this diverse research landscape and maximizing the collective impact of statistical contributions \citep{donoho2024data}.

While we have argued for the utility and necessity of statistical contributions to LLM development and application, we have not addressed the matter of timing. It remains uncertain whether generative AI, particularly LLMs, will lead to artificial general intelligence. However, it appears highly probable that these technologies will constitute a lasting and significant component of the future AI landscape. This presents a significant and timely opportunity for the statistics community \citep{lin2025statistics}. However, the risk of missing the chance to shape these AI technologies might arise if our community delays active engagement, since researchers in other fields, such as computer science---where younger generations often receive substantial statistical training---may develop these solutions on their own. While these contributions would still leverage statistical ideas, they might adopt a different flavor or lack the rigor that principled statistical approaches could provide. Waiting for the field of LLMs to ``stabilize'' or for problems to become ``well-defined'' risks allowing non-statistical or less-statistically grounded methodologies to occupy domains where principled statistical approaches would be more appropriate. However, the potential ``equilibrium'' is not necessarily \textit{unique}. Principled statistical approaches, if they come late, might not necessarily replace less-statistically grounded approaches that arrive earlier, especially in a field like LLMs involving significant engineering, scientific, and business considerations. Therefore, it is crucial for statisticians to be proactive to ensure that the development and application of LLMs benefit fully from the depth and rigor of statistical science.

\section*{Acknowledgments}

I am grateful to the associate editor and two referees for their constructive comments, which improved the presentation of this paper. I also thank Xiang Li for comments on an earlier version of the manuscript. This work was supported in part by NSF DMS-2310679, a Meta Faculty Research Award, and Wharton AI for Business.

%\clearpage
%{\small
\bibliographystyle{abbrvnat}
\bibliography{ref}
%}

\end{document}